\documentclass[12pt]{article}

\newcommand{\rdij}{{\tilde \rho}_{ij}}
\newcommand{\rij}{\rho_{ij}}

\newcommand{\be}{\begin{equation}}
\newcommand{\ee}{\end{equation}}
\newcommand{\beqs}{\begin{eqnarray}}
\newcommand{\eeqs}{\end{eqnarray}}

\expandafter\ifx\csname mathbbm\endcsname\relax

\else

\fi
\begin{document}

\begin{titlepage}

\begin{flushleft}  
\end{flushleft}
\vspace*{3mm}
\begin{center}
{\LARGE {Correlation breakdown, copula credit default models and arbitrage}\\}
\vspace*{12mm}
\large {Rodanthy Tzani\footnote{Present address: Bank Supervision Group, Federal Reserve
Bank of New York}} \\
\vspace*{1mm}
\large
{\em ACA Financial Guaranty Corp., New York}\\
\vspace{6mm}
\large{Alexios P. Polychronakos}\\
\vspace*{1mm}
{\em Physics Department, City College of the CUNY, New York}\\
\vspace{6mm}
\small rtzani@aca.com, alexios@sci.ccny.cuny.edu \/\\

\vspace*{15mm}
\end{center}

\begin{abstract}{

The recent `correlation breakdown' in the modeling of credit default swaps, in which model correlations
had to exceed 100\% in order to reproduce market prices of supersenior tranches, is analyzed and argued
to be a fundamental market inconsistency rather than an inadequacy of the specific model. As a consequence,
markets under such conditions are exposed to the possibility of arbitrage. The general construction of
arbitrage portfolios under specific conditions is presented.}

\end{abstract}

\end{titlepage}

\section{Introduction}

The high volatility in the market of credit derivatives, precipitated by the recent mortgage crisis,
has elevated the question of proper modeling and valuation of such products into one of the most
pressing and interesting problems in finance. Its relevance spans the spectrum from hedging to speculation
in the credit swap market.

The prevailing model for analyzing and valuing such products is the Gaussian copula model, otherwise known
as ``correlation model", loosely based upon Merton's theory. In this model, the problem of producing
or simulating
the defaults of a pool of entities (``names") under given individual default probabilities $p_i$ over a time
horizon and corresponding default
correlations $\rdij$ is addressed by subjugating the default events to the evolution of underlying normal
correlated variables, representing the ``assets" of each name. The default statistics produced this way
translate into a loss distribution for the portfolio containing these names, which, in turn, allows
the valuation of tranched credit default swaps referring to this pool. Variants of this model are used
by both market participants and rating agencies.

The market upheaval related to precipitous changes in the values of such default swaps has exposed this process as basically flawed. The reasons could be
manifold but they can be lumped into the following generic causes:

\noindent
$\bullet$ The model is unrealistic, inadequate or insufficiently sophisticated

\noindent
$\bullet$ The market has overreacted and created valuations inconsistent with the model as well
as reality

\noindent
$\bullet$ A combination of the above

\noindent
The third possibility, being the most generic, is the most likely. Still, even a partial assignment of blame
to the market must be fully substantiated. Eventually it is market prices that matter in finance and any model
is only as successful as its ability to describe the market.

A common instance in which markets can be assumed to misbehave is when market prices are way out of whack with fundamentals.
We have many such examples in early and recent history, but the problem is that it is hard to demonstrate
the discrepancy in present time; it is only {\it a posteriori} that such deviations from reality become
documented and (usually, painfully) obvious.

The situation in which markets misbehave in a presently detectable way is when they develop internal inconsistencies,
ranging from the plain and obvious to the subtle and convoluted. Such inconsistencies (or inefficiencies)
should be demonstrable in an unambiguous way through the analysis and comparison of market data themselves.
They also should lead to self-correcting effects in terms of opportunities for arbitrage. Transaction costs,
liquidity and other causes may frustrate the actual exploitation of these opportunities, but
still the identification of such situations as a diagnostic of market imbalance is a very interesting issue.

In this paper we focus on the recent `correlation breakdown' phenomenon of early 2008,
in which model correlations
above 100\% were invoked to explain market prices, as an obvious case study of possible inconsistencies.
Omitting many details, we shall show that, indeed, this situation points to market detuning and presents
identifiable arbitrage openings. Our analysis by no means absolves the present correlation model -in fact,
we consider it imperative that the model be refined or further developed. 
It does, however, shed some light on the
workings of the market and identifies one aspect in which the adoption of a better model may not be the main issue. 

As an outline of the results, the main points demonstrated in this paper are:

\noindent
$\bullet$ We recall the existence of upper limits for the values of the elements of the
default correlation matrix $\rdij$ for given default probabilities

\noindent
$\bullet$ We establish the uniqueness of the default process (joint default probabilities)
when the above limit is saturated

\noindent
$\bullet$ We demonstrate that the Gaussian copula model (and any other copula model)
produces the above unique process when asset correlations take their maximum value
of 100\%

\noindent
$\bullet$ We point out that situations where asset correlations higher than 100\% are needed
to reproduce market prices of tranches reveal fundamental inconsistencies in market prices

\noindent
$\bullet$ We construct appropriate arbitrage portfolios that exploit the above inconsistencies
by locking in riskless profits

\section{Defaults under maximal correlation conditions}

Consider a set of names with individual default probabilities at the end of a specific time horizon
$p_i$. Their state of default is represented by a set of default indicators $I_i$ assuming the value
1 if name $i$ has defaulted and 0 otherwise. The default correlations $\rdij$ are defined as
\be
\rdij = \frac{ P(I_i = 1 , I_j = 1 ) - p_i p_j}{\sqrt{p_i q_i p_j q_j}}
\ee
where $q_i \equiv 1-p_i$ is the probability of non-default of name $i$.

The $\rdij$ essentially determine the joint default probabilities of any two names, $P(I_i,I_j)$.
Indeed, there are four such probabilities, corresponding to the four possible default
scenarios of names $i$ and $j$; they must reproduce $\rdij$, $p_i$ and $p_j$ and sum to one; that is,
\beqs
P(1,1) - p_i p_j &=&\rdij \sqrt{p_i q_i p_j q_j} \cr
P(1,0)+P(1,1)&=&p_i \cr
P(0,1)+P(1,1)&=&p_j \cr
P(0,0)+P(0,1)+P(1,0)+P(1,1)&=&1
\eeqs
These are enough relations to fully determine the $P(I_i,I_j)$ as
\beqs
P(0,0)&=&q_i q_j + \rdij \sqrt{p_i q_i p_j q_j} \cr
P(0,1)&=&q_i p_j - \rdij \sqrt{p_i q_i p_j q_j} \cr
P(1,0)&=&p_i q_j - \rdij \sqrt{p_i q_i p_j q_j} \cr
P(1,1)&=&p_i p_j + \rdij \sqrt{p_i q_i p_j q_j} 
\eeqs

The above probabilities must be non-negative. (They must also not exceed l, but this is guaranteed
by non-negativity and the fact that they add to 1.) This implies constraints for $\rdij$. The first
and last relations above imply a negative lower bound. Since correlations are generally assumed positive,
this is not a useful or restrictive constraint. The middle two relations, however, together with
$P(0,1) \ge 0$  and $P(1,0) \ge 0$, imply
\be
\rdij \le min \left( \sqrt{\frac{p_i q_j}{q_i p_j}} , \sqrt{\frac{q_i p_j}{p_i q_j}} \right)
\label{cons}
\ee
This is a consistency condition that each element of any default (binary) correlation matrix must satisfy.

It is worth pointing out that the above condition is distinct from the condition of positivity
of the eigenvalues of $\rdij$. There are positive definite correlation matrices violating the
above condition; conversely, there are non-positive definite matrices satisfying the above condition.
Consequently, satisfaction of the above condition does not guarantee that $\rdij$ is a possible
default correlation matrix, meaning that well-defined joint default
probabilities leading to the above correlations may not exist.

Interestingly, however, when the above conditions are {\bf saturated} (that is, they hold as equalities)
for all matrix elements of $\rdij$, the corresponding joint default probabilities do exist and are
{\bf unique}. To demonstrate the existence, assume that we order the names in terms of increasing
default probability; that is,
\be
p_1 \le p_2 \le \dots \le p_N
\label{ord}
\ee
The default correlation matrix that saturates the condition (\ref{cons}) reads
\be
\rdij = \sqrt{\frac{p_i q_j}{q_i p_j}} ~,~~~ i<j
\label{crit}
\ee
Then the joint default probabilities that realize the above correlations are of a `ladder' type:
\be
P(I_1 = \dots =I_n = 0, I_{n+1} = \dots =I_N = 1) = p_{n+1} - p_n
\label{Pnn}
\ee
all the rest being zero. (To make the above formula valid for $n=0$ and $n=N$ we define 
$p_0 =0$ and $p_{N+1} = 1$.) In other words, names default in a hierarchical order, each name
defaulting only if all names with higher default probability also default.

As a simple explicit example, consider a portfolio with 5 names of
default probabilities
\be
p_1 = 0.6\% ~,~~p_2 = p_3 = 1\% ~,~~ p_4 = 1.2\% ~,~~ p_5 = 4\%
\ee
Then the only default scenarios with non-zero probability are:

\noindent
a) No names default, with probability $1-p_5 = 96\%$

\noindent
b) Only name 5 defaults, with probability $p_5 - p_4 = 3.8\%$

\noindent
c) Only names 4 and 5, default, with probability $p_4 - p_3 = 0.2\%$

\noindent
d) Only names 2,3,4 and 5 default, with probability $p_2 - p_1 = 0.4\%$

\noindent
e) All names default, with probability $p_1 = 0.6\%$

\noindent
Note that names 2 and 3 always default in tandem, since $p_2 - p_3 =0$,
and if any name defaults the whole set of names riskier than this name 
also default.

The default process defined in (\ref{Pnn}) reproduces the correct single-name default probabilities $P(I_n=1)=p_n$ and
default correlations (\ref{crit}). These facts, and the fact that it is the unique process that reproduces the above default probabilities and
correlations, are easy to prove and we present the derivations in Appendix A.

The lesson derived from this section is that for any given set of default probabilities there is
an absolute maximum in the default correlations and a unique default pattern that realizes this maximum.
Any model that can reproduce the above pattern with appropriate imput parameters has achieved the maximum
possible correlations.

\section{Maximally correlated copula models}

The model most widely used to produce default events with given 
$\{ p_i , \rdij \}$ is the Gaussian
copula model. The basic steps of a simulation based on this model can be outlined as

\noindent
$\bullet$ Assign to each name a continuous variable $X_i$ (``asset") whose value at the end of the time period 
of interest is normally distributed

\noindent
$\bullet$ Assume that a name will default if the value of its asset variable at the end of the period has 
fallen below a specific minimal value $c_i$ (``threshold")

\noindent
$\bullet$ Render the asset variables correlated with a given correlation matrix $\rij$

\noindent
$\bullet$ Adjust the parameters $c_i$ and $\rij$ such that the above model reproduce the desired results.

Market models adjust the above parameters so that they reproduce the market prices of a given set of credit products.
Rating agencies, on the other hand, use the default data $p_i$ and $\rdij$ as the fundamental inputs.
Determining the  Gaussian inputs $\{c_i , \rij \}$ in terms of the binomial inputs $\{p_i , \rdij \}$ is  
achieved using numerical or approximate analytic formulae for the Gaussian
probabilities  $P( X_i < c_i ) = p_i$ and $P ( X_i < c_i , X_j < c_j ) = p_i p_j +\rdij \sqrt{p_i q_i p_j q_j}$.

The ``correlation model" often used in the market in order to define
base correlations is a further simplified version of the Gaussian
copula model with the additional assumption of homogeneity: all names
have equal notional amounts, the same default probability (deduced from their
average spread) and `flat' asset correlations, that is, 
$\rij = \rho =$ constant (a `single-factor' model). This reduces the
number of parameters and compromises the model, making it a less
realistic description of the default process. In what follows we shall
use the more general (non-homogeneous) Gaussian copula model. As we
shall argue, the appearance of oversized ($>$100\%) correlations in the
homogeneous model would also imply correlation breakdown in the more
general non-homogeneous model and thus also the emergence of
arbitrage situations.

The asset correlation matrix $\rij$, apart from being positive definite, is unrestricted. The maximally
correlated model is, then, achieved by choosing all asset correlations equal to 100\%, that is, $\rij =1$.
Under such conditions, the Gaussian copula model reproduces the maximal
correlation default model of the previous section, with default
correlations as in (\ref{crit}) and joint default probabilities as
in (\ref{Pnn}).

To prove this fact it is enough to notice that fully (100\%) correlated Gaussian variables are essentially
the same variable; that is, all asset variables can be set to a unique Gaussian variable $Z$: $X_i =Z$.
For names ordered according to their default probabilities, as in (\ref{ord}), the thresholds are 
similarly ordered:
\be
c_1 \le c_2 \le \dots \le c_N
\ee
For a value of $Z$ such that $c_n < Z < c_{n+1}$, the names $n+1 , \dots , N$ default (since $X_i = Z < c_i$
for such names) while the names $1, \dots n$ do not default ($X_i = Z > c_i$). This reproduces the `ladder'
default pattern of the previous section. Further, since
\be
P(c_n < Z < c_{n+1}) = P(Z < c_{n+1}) - P(Z < c_n) = p_{n+1} - p_n
\ee
these probabilities are as in (\ref{Pnn}). So we recover the exact default probabilities of the (unique)
maximally correlated default model.

It should be clear that {\bf any} copula model using continuous, correlated variables will also reproduce
the above maximally correlated default model upon choosing asset correlations 100\%. Indeed, any set of
fully correlated continuous variables will collapse into a unique variable $Z$. Since in the above analysis
we never used the specific form of the Gaussian cumulative distribution $P(Z<c)$, it goes through for any
copula model.\footnote{The above discussion shows that at maximum correlation
all simulation models become equivalent and can be realized in the
following way: Produce a random number $X$ uniformly distributed
between 0 and 1. Then all names with default probability bigger than
$X$ default, while all names with default probability less than $X$ do not. 
This provides a simple and expedient way to simulate this system.}

The lesson derived here is that all copula models with the same default
probabilities $p_i$ will produce identical joint default probabilities for
maximally correlated defaults, and therefore identical results for the
loss distribution of the reference portfolio and tranche pricing, provided
that all other assumptions, i.e., recoveries, are the same.
In general, different models may be more or less successful, accurate or realistic. For market conditions
calling for highly (in fact, maximally) correlated defaults, nevertheless, they are all identical and produce
the unique consistent default pattern.

\section{Market prices and correlation breakdown}

In considering the credit default process of a set of names, we may adopt two distinct poits of view on
the nature of the joint default probabilities:

\noindent
$\bullet$ They represent objectively defined fundamental probabilities that describe the actual default
process. They can be deduced from historical data or by best-estimate analysis based on current fundamentals.

\noindent
$\bullet$ They represent market-implied probabilities that reproduce market prices through risk-neutral
valuation. They can be deduced by tuning model parameters to match the valuation of market-priced entities.

It is the second point of view that is relevant to the present discussion. Implied probabilities have been
used widely in finance: they are, e.g., the `Martingale' probabilities of options pricing under complete and
efficient market assumptions. More importantly, they are the basis for calibrating the ``correlation model"
that has become the standard in pricing bespoke tranches of credit default swaps.
They are the probabilities that would enter any discussion of price compatibility and arbitrage.

It could be argued that, under this second point of view, implied probabilities are not `true' probabilities
in the mathematical sense and thus may not need to share all the properties and satisfy all the constraints
of probability theory. Indeed, implied tranche correlations (or even the better-behaved base correlations)
of the correlation model, in order to match market prices are varied with impunity over detachment points even as they
refer to the same pool of names, a fundamentally inconsistent procedure in the probabilistic sense.

It is important, nevertheless, to realize that implied probabilities do behave as regular probabilities and
that model-independent deviations from regular behavior signal market pricing inconsistencies and create, in principle, the
possibility for arbitrage trading. This is the topic of the present section.

The strongest signal of such a deviation from regular behavior was the `correlation breakdown' of early 2008, which required a flat asset correlation of more that 100\% in the homogeneous copula model to reproduce the market prices of supersenior tranches.\footnote{In principle, the model works with base correlations that refer
to equity tranches of given detachmet points. Because of equity-supersenior parity, however, the two are
directly related: the sum of an equity tranche and a supersenior tranche that share the same detachment-attachment
point is the full portfolio, whose value is correlation-blind; therefore the implied correlations of the
two tranches are equal.}
Note that, for the homogeneous model, asset correlations of 100\%
also imply default correlations of 100\%.

Correlations of more than 100\% in the simulation model cannot exist. To see what they mean in practice,
consider the value of a supersenior tranche obtained through risk-neutral valuation as the expectation of
the value of the tranche at maturity. This is, in general, an increasing function of the default correlations,
and thus also of the (flat) asset correlations. A semi-analytic formula yielding the value of the tranche as
a function of the asset correlation can be established (at least in the
large-pool homogeneous model). If at the maximum asset correlation of 100\% we still
obtain a value below the market price of the tranche, we can drive the
correlation higher than 100\%, beyond the range of validity of the
formula, in order to reach the market price. 
This is the correlation breakdown effect.

What the above means is that, even for the maximum asset correlation of 100\%, the Gaussian copula model fails
to produce a tranche price as high as its market value. If this is the
case for the homogeneous model, it will also happen for the more
general non-homogeneous model. Indeed, it can be seen that, for 100\%
asset correlations, the homogeneous model will give a more widely
spread portfolio loss distribution than the non-homogeneous model,
increasing the expected loss of supersenior tranches and thus the
fair value of their credit default swap. If this is still below the
market value, so will be the one calculated from the non-homogeneous
model.

The discrepancy could still be attributed to a failure of the specific model. We
know, however, from the analysis of the previous section that this is not the case: the Gaussian model (as
well as any other model) produces the unique default process with the highest possible default correlations
under given individual name default rates. If even this model underestimates the value of the tranche, this
means that the market's pricing of the tranche is unreasonably high and inconsistent with the underlying
name default rates.

The individual name default rates are `priced' by the market in terms of credit
default swaps for the corresponding individual names. The existence of an independent market product (the supersenior
tranche) with a valuation fundamentally inconsistent with the prices of its underlying components should signal the
existence of arbitrage opportunities. Indeed, this is the case as will be demonstrated in the next section.

\section{The arbitrage portfolio}

In considering the question of arbitrage we shall make the usual assumptions of friction-free trading:
enough liquidity to go long or short any amount in all available products with negligible bid-ask spread and
transaction costs.

We shall work with a pool of $N$ defaultable names. Credit default swaps (CDS) are available for each name
as well as tranches of a reference portfolio containing a (percent) notional amount $N_i$ of each name.
To reduce the issue to its bare bones, we assume a single-period
situation in which all premiums are paid up-front and any losses are incurred at the end. We shall also
assume fixed (deterministic) recovery rates, different for each name. This is a simplifying assumption,
in line with most standard model simulations; stochastic, correlated recoveries clearly would require
a different analysis. Finally, we assume zero risk-free interest rates (a harmless assumption allowing us
to dispense with present-value conversion factors).

The fair present value of a credit product $C$ is its expected value at maturity. Denoting with
$\{ I_i \}$ the collective state of default of the names at maturity, represented by their default
indices, $P(\{ I_i \})$ the probability of this default scenario and $V(C \{ I_i \})$ the value of the
credit product at maturity for the given default scenario, the risk-neutral fair present value of the product is
\be
V(C) = \sum_{\{I_i =0,1\}} P(\{ I_i \}) \, V(C\{ I_i \})
\label{V}
\ee
For a credit default swap, $V(C\{ I_i \})$ is the payment that the seller of protection must make at
maturity to cover losses due to the default scenario $\{ I_i \}$, and thus $V(C)$ is the fair premium that
the seller of protection must receive up-front.
$P(\{ I_i \})$ are market-implied probabilities that reproduce the market prices $V(C)$ for all available
CDSs.

Under the above assumptions, the market value of a CDS on \$1 notional of name $i$, denoted $C_i$,
assuming a deterministic recovery $r_i$ and loss given default $\ell_i = 1-r_i$, is
\be
V(C_i) = \ell_i \, p_i
\label{VB}
\ee
In this sense the individual CDS $C_i$ `prices' the implied probability of default $p_i$.
Similarly, the market value of a CDS for a supersenior tranche with attachment point $A$, denoted $S[A]$, is
\be
V(S[A]) = \sum_{\{I_i =0,1\}} P(\{ I_i \}) \, \left[\sum_{i=1}^N N_i \ell_i I_i \, -A \right]_+
\ee
where $[x]_+ = max(x,0)$ is defined as $x$ for $x>0$ and $0$ otherwise. (Note that a
supersenior CDS with attachment point higher than $\sum_i N_i \ell_i$ is never hit and thus has
vanishing fair value. This is a known problem of fixed recovery rates.)

We are now ready to present an arbitrage scenario. Assume the names are arranged in order of increasing
default probability, as in previous sections, and that the attachment point $A$ of the supersenior
tranche lies within the range:
\be
A = \sum_{i=n+1}^N N_i \ell_i \, + \epsilon N_n \ell_n ~,~~~ 0< \epsilon \le 1
\label{range}\ee
Then without further elaboration we propose the {\bf arbitrage portfolio}:

\noindent
$1.$ We sell protection on the supersenior tranche (go long on one unit of $S[A]$)

\noindent
$2.$ We buy protection on $N_i$ units of name $i$, for all names $i=1, \dots n-1$ (go short on
$N_i$ units of $C_i$)

\noindent
$3.$ We buy protection on $(1-\epsilon)N_n$ units of name $n$ (go short on $(1-\epsilon) N_n$ unit of $C_n$)

We claim that:

\noindent
$\bullet$ The portfolio $P_{arb}$ will {\bf always have a non-negative value at maturity}, no matter what the actual
default scenario

\noindent
$\bullet$ Under market conditions of `correlation breakdown', the portfolio $P_{arb}$ will have a
{\bf negative initial value}

The above two statements imply that, under correlation breakdown, the portfolio $P_{arb}$ will always increase
in value irrespective of the actual defaults. Therefore, it will afford a riskless profit, realizing an
arbitrage opportunity in the market. The proof of the above statements is given in Appendix B.

We should stress that the above portfolio is not fluctuation-free: its value at maturity will vary according
to the realized default scenario. In this sense it is different than the arbitrage portfolios in Cox-Ross or
Black-Sholes options pricing theory, which assume a deterministic value at maturity. Nonetheless, it
is a clear case of arbitrage as it represents a `can't-lose' gamble.

The structure of the arbitrage portfolio is remarkably simple: we buy protection for specific amounts of the
$n$ {\bf least} risky names and sell protection for the supersenior tranche starting at $A$. This is somewhat
counterintuitive, as the tendency would be to buy protection on the {\bf riskiest} assets in the reference
portfolio. Under correlation breakdown conditions
the market overvalues the protection of the supersenior tranche. As a result, the premium
collected from this tranche is enough to cover the premiums paid for the protection of individual names, while at
maturity no net losses can possibly be incurred. Appendix B makes these statements exact.

The dressed-down situation that we considered above has shed many of the features of realistic traded
credit default swaps.
As it refers to a single-period swap with up-front premium payments and end-of-period default payments,
it does not deal with issues of default timing and corresponding reduction of received premiums, resulting
in a simpler valuation. (This is what makes formulae (\ref{V}) and (\ref{VB}) valid.) Also, the conditions for liquidity and small bid-ask spreads may not be fully met in available credit products. 

The assumption
of deterministic recovery, although standard in most models, does not reflect the reality of defaults.
Incorporating stochastic, correlated recoveries would require a different 
analysis.\footnote{Stochastic, uncorrelated defaults would have only a marginal impact as the central-limit theorem 
would wash out their stochastic nature for a large number of defaults.} To produce a true arbitrage portfolio in that
case we would need to cover the possibility that actual recoveries are small. This is achieved by
putting all $\ell_i =1$ in formula (\ref{range}). This will increase the value of $n$ and will result in buying
protection on a larger number of the least risky assets $1, \dots ,n$, thus making the portfolio more expensive.
The market price for the supersenior tranche that would guarantee arbitrage would need to exceed the premiums paid
for this expanded set of single-name CDSs.

In general, the above price would be
higher than the model price for 100\% asset correlations and fixed
recovery assumptions. The model that reproduces this price with 100\% correlations must assume 0\% recoveries for names $n, \dots ,N$ (recoveries of the least risky assets $1, \dots ,n-1$ need not be lowered, as they have no impact on the fact that the arbitrage portfolio suffers no net losses at maturity.) Since reported model
correlations exceeding 100\% assumed, in general, fixed nonzero recoveries, the existence of an arbitrage portfolio with mathematically
zero probability of loss in this more general situation is unclear.
The above analysis, nevertheless, serves as an indication that market prices during the recent correlation
breakdown have moved
beyond reasonable bounds, at least for senior tranches with short maturity.

\section{Conclusions}

We have identified a situation in which the market has overreacted to credit events and appears to have priced
some supersenior CDS tranches in a fundamentally inconsistent way. Regardless of whether the resulting
arbitrage opportunities were actually realized and contributed to the normalization of the market, the fact
remains that market data cannot always be taken at face value when pricing credit derivatives or other
financial products.

The above stresses the importance of fundamental analysis, modeling and risk management. A good model that
consistently takes into account fundamental principles and properly incorporates the underlying market forces,
rather than trying to slavishly mimic and match market data, is badly needed. (A ``first-principles" rather
than ``phenomenological" model, in physics lingo.) Market participants should exercise proper discipline and
operate competent and independent risk management teams, otherwise risking
to create and fall into their own ``negative vacuum" bubbles (another physics allusion). Finally, rating
agencies should return to their mission and be more proactive and original,
thus reinforcing their credibility and offering valuable market checks and balances.
In this context, independent valuation companies with strong quantitative, analytical and modeling
teams play an increasingly important role and fill a crucial market void.

\vskip0.5cm \noindent \underline{Achnowledgements:} We would like to thank Joseph Pimbley for a critical reading 
of the manuscript and for many insightful comments and remarks that helped make this a better paper.


\vskip 0.2in
{\centerline {\large {\bf APPENDIX A}}}

\vskip 0.3in

The condition $\rdij = \sqrt{p_i q_j / q_i p_j}$ for any two names $i$ and $j$ (with $p_i \le p_j$) implies
that $P(I_i = 1 , I_j = 0) = 0$. This two-name marginal probability is expressed as
\be
P(I_i = 1 , I_j = 0) = \sum_{\{I_k = 0,1\}} P(I_1, \dots I_{i-1}, 1 , I_{i+1} , \dots , I_{j-1}, 0 , I_{j+1} ,
\dots I_N)
\ee 
The above sum of non-negative terms vanishes only if all individual terms are zero. We conclude that all
probabilities $P(I_1 , \dots , I_N)$ with even a single pair $i<j$ defaulting in the `wrong' order are zero.
The only possible nonzero probabilities are of the form
\be
P_{(n)} = P(0, \dots , 0,1, \dots , 1) ~,~~~ n=0,\dots,N
\ee
with the first $n$ names not defaulting and the remaining $N-n$ names defaulting. This proves the `ladder'
(hierarchical) pattern of defaults at maximal correlations. On the other hand the single-name default
probabilities are expressed as
\be
p_i = \sum_{\{I_k =0,1\}} P (I_1, \dots I_{i-1}, 1 , I_{i+1} , \dots , I_N ) = \sum_{n=0}^{i-1} P_{(n)}
\ee
This recursion relation, under the initial condition $p_1 = P_{(0)}$, has the unique solution
\be
P_{(n)} = p_{n+1} - p_n
\ee
as in (\ref{Pnn}), proving the uniqueness of the process. Further, from the above form of probabilities
we have
\be
P(I_i = 1 , I_j = 1) = \sum_{n=0}^{i-1} P_{(n)} = p_i ~,~~~ i<j
\ee
which gives $\rdij = \sqrt{p_i q_j / q_i p_j}$ as expected.

\vfill\eject

{\centerline {\large {\bf APPENDIX B}}}


\vskip 0.3in

We shall first show that the portfolio $P_{arb}$ will always have a non-negative value at maturity.

If any of the first $n-1$ names default the portfolio suffers no loss: we shall receive
a payment equal to $N_i \ell_i$ for each $C_i$ in $P_{arb}$, while the subordination of $S[A]$ will decrease by
the same amount, so $S[A]$ will suffer a loss of at most this same amount, if any at all.
Therefore, defaults of these names are harmless or even beneficial for the portfolio.
We need concentrate only on the defaults of the remaining $N-n+1$ names.

Out of all default scenarios, the most potentially damaging for portfolio $P_{arb}$ are the ones in which either 
the last $N-n$ names have defaulted or the last $N-n+1$ names have defaulted.
In the first case the reference portfolio has suffered a loss
\be
L = \sum_{i=n+1}^N N_i \ell_i < A
\ee
and therefore the tranche with attachment point $A$ is not hit. We neither receive nor make any payments,
for a net value of zero.

In the second case the reference portfolio has suffered a loss
\be
L = \sum_{i=n}^N N_i \ell_i = \sum_{i=n+1}^N N_i + \epsilon N_n \ell_n + (1-\epsilon) N_n \ell_n
= A + (1-\epsilon) N_n \ell_n
\ee
The tranche with attachment $A$ has been hit by an amount $(1-\epsilon) N_n \ell_n$ and $S[A]$ has incurred
this amount as a loss. On the other hand, the $(1-\epsilon) N_n$ units of $C_n$ in the portfolio
receive a payment of $(1-\epsilon) N_n \ell_n$ due to the default of name $n$. The portfolio
breaks even, again for a net value of zero.

Overall, we see that in any default scenario the portfolio will either make a profit or at least break even.
A profit will be made if some of the first $n$ names default without all the remaining ones defaulting.

We shall now show that the initial value of the portfolio is negative if correlation breakdown occurs.

The value of the individual CDSs included in $P_{arb}$ depends only on $p_i$ and is independent of correlations.
The value of the supesenior tranche $S[A]$, on the other hand, increases with increasing correlation. Therefore
as correlations increase the value of the portfolio $P_{arb}$ drops.

For the limiting consistent case of 100\% asset correlations it is easy to calculate the value of $P_{arb}$.
As explained
in this paper, in that limit the default pattern is unique and the only default scenarios with nonzero probability
are the ones where names default hierarchically, riskiest (last) ones first. If fewer than the last $N-n+1$ names
default the portfolio $P_{arb}$ incurs no losses and receives no payments. If exactly $N-n+1$ names default the portfolio
again breaks even, as explained before. If all $N-n+1$ riskiest names plus a number of the remaining ones default
the supersenior tranche incurs extra losses, which are exactly balanced by the protection payments received from
the single-name CDSs included in $P_{arb}$. Overall, we see that the portfolio breaks even in each of the scenarios that
have nonzero probability. We conclude that the present value of the portfolio for maximal consistent correlations
is zero.

If the market prices supersenior tranches higher than the fair price for 100\% asset correlations we have the
correlation breakdown situation. In that case the single-name CDS part of $P_{arb}$ has the same value but the value
of $S[A]$ is higher, and thus the value of $P_{arb}$ is lower. We have therefore proved that under correlation breakdown
market conditions the portfolio $P_{arb}$ has negative present value.

We should note that it is possible to show rigorously the negative present value of $P_{arb}$ without reference to
the fact that it is a decreasing function of correlations. We found it preferable, however, to use the familiar
fact of increasing supersenior value with increasing correlations to present a shorter and perhaps more intuitive
derivation of the negative present value property of $P_{arb}$.

The two properties of $P_{arb}$ proven in this appendix establish its arbitrage property of securing riskless profits.

\end{document}